\makeatletter
\@ifundefined{@parse@version@dash}{%
\def\@parse@version#1{\@parse@version@0#1}
\def\@parse@version@#1/#2/#3#4#5\@nil{%
\@parse@version@dash#1-#2-#3#4\@nil}
\def\@parse@version@dash#1-#2-#3#4#5\@nil{%
  \if\relax#2\relax\else#1\fi#2#3#4 }
}{}
\makeatother

\documentclass[%
 reprint,
 amsmath,amssymb,
 aps,
]{revtex4-1}
\usepackage{amsthm}
\usepackage{dcolumn}
\usepackage{bm}

\usepackage[colorlinks]{hyperref}
\makeatletter

\newcommand{\Rmnum}[1]{\expandafter\@slowromancap\romannumeral #1@}
\makeatother
\begin{document}
\title{Quantifying coherence in terms of Fisher information}
\author{Deng-hui Yu$^{1}$}

\author{Chang-shui Yu$^{2}$}
\email{ycs@dlut.edu.cn}

\affiliation{$^1$School of Information Engineering, Zhejiang Ocean University, Zhoushan
316022, China\\
$^2$School of Physics, Dalian University of Technology, Dalian
116024, China}

\date{\today }

\begin{abstract}
In quantum metrology,  the parameter estimation accuracy is bounded by quantum Fisher information.
In this paper, we present coherence measures in terms of (quantum) Fisher information by directly considering the post-selective non-unitary parametrization process.
This coherence measure demonstrates the apparent operational meaning by the exact connection between coherence and parameter estimation accuracy.
We also discuss the distinction between our coherence measure and the quantum Fisher information subject to unitary parametrization. The analytic coherence measure is given for qubit states.
\end{abstract}

\maketitle

\section{introduction}

Quantum coherence, as a fundamental feature in quantum physics, attracts a lot of attention in recent years.
Many works have investigated the role of coherence in quantum optics \cite{mandel1995optical,op1,op2,op3}, quantum thermodynamics \cite{Mitchison_2015,PhysRevLett.113.150402,Korzekwa_2016}, quantum phase transitions \cite{PhysRevB.90.104431}, quantum biology \cite{PhysRevLett.106.040503,Lloyd_2011}, and quantum information science \cite{doi:10.1126/science.1104149,Giovannetti2011,PhysRevLett.96.010401,PhysRevA.95.032307,PhysRevA.52.3489,doi:10.1098/rspa.1992.0167,
PhysRevA.93.012111,PhysRevLett.79.325}.
These researches not only promote the development of related applications but also the development of the resource theory of coherence \cite{RevModPhys.89.041003,PhysRevLett.113.140401}, where coherence is treated as a physical resource under some limited conditions.
Benefiting from  the operational view and axiomatic approach, one can quantify coherence in a rigorous manner,
study the transformation of coherence, and  reveal the
connection between coherence with other fundamental quantum features \cite{RevModPhys.81.865,PhysRevLett.115.020403,PhysRevLett.117.020402,PhysRevLett.121.220401,
PhysRevA.96.032316,PhysRevLett.88.017901,Henderson_2001,PhysRevLett.116.160407,RevModPhys.86.419,PhysRevA.100.022310,PhysRevA.95.010301,PhysRevA.80.022324}.
In particular,   some coherence measures contain obvious operational meanings, which provide us with a way to understand (interpret) coherence from the viewpoint of quantum information processes (QIP) and find out the potential relation between coherence with some characteristics in QIP \cite{PhysRevLett.116.120404,PhysRevA.92.022124,PhysRevLett.121.010401,PhysRevLett.120.070403,PhysRevLett.116.150502,PhysRevA.95.042337,PhysRevA.101.062114}.

It is shown that the coherence of the probing state in many quantum metrology processes is often a key ingredient \cite{Giovannetti2011,doi:10.1126/science.1104149,PhysRevLett.96.010401}.
For instance, in the usual phase estimation for parameter $\theta$ with unitary parametrization $\mathcal{U}_\theta(\cdot)=e^{-i\theta H}(\cdot)e^{i\theta H}$,
coherence with regard to the eigenvectors of Hermitian operator $H$ is necessary.
Furthermore,
the optimal estimation accuracy of an unknown parameter could be obtained by the state with maximal coherence in the sequential protocol \cite{PhysRevLett.96.010401}.
The estimation accuracy is bounded by quantum Fisher information (QFI), a crucial ingredient in quantum metrology \cite{van1998cambridge,hayashi2006quantum,PhysRevLett.72.3439}.
A simple calculation can show that QFI subject to unitary parametrization $\mathcal{U}_\theta(\cdot)$ in the qubit case \cite{PhysRevA.87.022337} is monotonous with some coherence measures (such as $l_1$ norm coherence).
Many works have investigated the relation between quantum coherence with Fisher information (FI) and QFI \cite{PhysRevA.98.022306,PhysRevA.96.022136,dephasing,u2,c1,PhysRevA.103.012401,c3,Feng2017,doi:10.1098/rspa.2017.0170,KWON20181594}.
Coherence within some \textit{particular settings} could be understood by QFI (or FI) \cite{dephasing,u2,doi:10.1098/rspa.2017.0170}.
Significantly, QFI in unitary parametrization is closely connected with unspeakable coherence \cite{u1,u2},
a special case of resource theory of asymmetry \cite{RevModPhys.79.555,Gour_2008,Marvian2014}.
In addition, based on QFI concerning the dephasing parameter, coherence measure has been given in the sense of strictly incoherent operations as free operations \cite{dephasing}.
However, up to now the estimation accuracy and FI (or QFI) have not been used to directly quantify quantum coherence in \textit{general} scenarios.
An intuitive challenge is that QFI with unitary parametrization $\mathcal{U}_\theta(\cdot)$ in the usual sense is not a coherence measure in the \textit{general} resource theory of coherence \cite{PhysRevLett.113.140401}.
For example,
2-dimensional maximally coherent states (MCS) could be obtained under incoherent operations from 3-dimensional MCS \cite{c1,PhysRevA.100.032313,PhysRevLett.116.120404},
but the QFI of the former is strictly larger than the latter, which directly violates the monotonicity of a good measure. Therefore, it is significant to find an appropriate parametrization process for establishing coherence measures and further investigating the role of coherence in quantum metrology.

In this paper, we successfully establish several equivalent coherence measures in the general resource theory of coherence by the FI (and QFI) subject to a type of non-unitary parametrization. Since the optimal estimation accuracy is bounded by FI which is asymptotically attained with maximum likelihood estimators \cite{van1998cambridge,hayashi2006quantum},
our measure naturally inherits the operational meaning of FI through the optimal estimation accuracy with non-unitary parametrization.
We also show that in the qubit case, our coherence measure can be equivalently understood through unitary parametrization and the analytic expression can be obtained.
Our coherence measure not only builds a direct relation between  coherence and parameter estimation accuracy (or FI) but also
sheds new light on the roles of the non-unitary parametrization process.
The remainder of this paper is organized as follows. In Sec. II, we first introduce the fundamental concepts of resource theory of coherence and our parametrization process, then present several main theorems to build the coherence measure based on FI.
In Sec. III, we give the analytic result of the coherence measure in the qubit case and discuss the equivalence with the unitary parametrization. Finally, we draw our conclusion in Sec. IV.

\section{Coherence in terms of QFI}
In this section, we'd like to first introduce the resource theory of coherence established mainly based on the incoherent (free) operations and incoherent (free) states \cite{PhysRevLett.113.140401}.  Considering the preferred basis $\{\left\vert n\right\rangle \}$, the incoherent state is defined by $
\varrho=\sum_nq_n|n\rangle\langle n|
$
with $\mathcal{I}$  denoting the set of incoherent states, and the incoherent operations (IO) with the Kraus representation $\{K_n: \sum_lK_l^\dagger K_l=\mathbb{I}\}$ is a special type of completely positive and trace-preserving (CPTP) map defined by
$
\frac{K_l\varrho K_l^\dagger}{\operatorname{tr}(K_l\varrho K_l^\dagger)}\in\mathcal{I}$ for $\varrho\in \mathcal{I}$.
 In this sense, a good coherence measure $C(\rho )$ for any state $\rho $ should satisfy

($A1$) {\textit{Non-Negativity}:} $C(\rho )\geq 0$ is saturated iff $\rho
\in \mathcal{I}$;

($A2$) {\textit{Monotonicity}:} $C(\mathcal{E}(\rho ))\leq C(\rho )$ for
any incoherent operation $\mathcal{E}(\cdot )$;

($A3$) {\textit{Strong Monotonicity}:} $\sum_{n}p_{n}C(K_{n}\rho
K_{n}^{\dagger }/p_{n})\leq C(\rho )$ for any IO $\{K_n\}$, with $p_{n}=\mathrm{Tr}[K_{n}\rho
K_{n}^{\dag }]$;

($A4$) {\textit{Convexity}:} $C(\rho )\leq \sum_{i}p_{i}C(\rho _{i})$ for
any $\rho =\sum_{i}p_{i}\rho _{i}$.

To present a valid coherence measure, we begin with the following parametrization process. Considering a state $\rho$ undergoing quantum channel $\mathcal{E}_\theta$ depending on parameter $\theta$, the unknown parameter could be estimated from measurements on $\mathcal{E}_\theta(\rho)$. Here we are interested in the ``free" parametrization processes $\mathcal{E}_\theta=\{E_x(\theta)\}$:
\begin{align}
&E_x(\theta)=\sum_nb_n^x(\theta)|g_x(n)\rangle\langle n|, \sum_xE_x(\theta)^\dagger E_x(\theta)=\mathbb{I},\label{IO}
\end{align}
where $\{|n\rangle\}$ is the preferred incoherent basis, and
$g_x(\cdot)$ is a map from an integer to another.

In order to focus on the role of coherence, we desire that within the parametrization process, the incoherent probe can not take effect on parameter estimation.
That is, the measurement outcomes $\mathcal{E}_\theta(\varrho)$ and $\{\varrho_x, p_x \}$ obtained from incoherent probe ($\varrho\in\mathcal{I}$) do not depend on parameter $\theta$, where $p_x=\operatorname{tr}[E_x(\theta)\varrho E_x(\theta)^\dagger]$ and $\varrho_x=E_x(\theta)\varrho E_x(\theta)^\dagger/p_x$. Thus $|b_n^x(\theta)|$ does not depend on parameter $\theta$, and $E_x(\theta)$ can be rewritten as
\begin{align}
E_x(\theta)=\sum_nc_n^xe^{ih_n^x(\theta)}|g_x(n)\rangle\langle n|,\label{dy}
\end{align}
where $c_n^x$ is parameter-independent, and $h_n^x$ is a real function. In fact, it is very similar to the case of the usual phase estimation $\mathcal{U}_\theta(\cdot)=e^{-i \theta H}(\cdot)e^{i\theta H}$ mentioned in Introduction.
One can find that the measurement outcomes of an incoherent probe in the phase estimation do not depend on parameter $\theta$, either. In addition, $\mathcal{U}_\theta$ can be expressed based on $e^{-iH\theta}=\sum_ne^{-ih_n\theta}|n\rangle\langle n|$
($h_n$ is eigenvalue of $H$), which is analogous to Eq. (\ref{dy}). In this sense, parametrization process $\mathcal{E}_\theta$ can be understood as a generalization of unitary phase estimation to non-unitary case.

In addition, we could restrict
$\partial_\theta h_n^x(\theta)\in[0,1]$, and the conclusion in a more general case could be derived from this case, the detailed discussion is shown in Appendix A. Based on the Stinespring dilation theorem, the operations could be implemented by a
controlled unitary operator and an operation swapping specified states. The details are shown in Appendix B.
All the operations of interest (operations in Eq. (\ref{dy}) with $\partial_\theta h_n^x(\theta)\in[0,1]$) comprise a set denoted by $G$.
One will find that IO satisfying $Rank\left [E_x(\theta)^\dagger E_x(\theta)\right]=1$ has particular interest in the paper, so we use $G_1$ to represent the IO set with this particular property.

If the post-selection is allowed, the IO $\mathcal{E}_\theta$ performed on a quantum state $\rho$ will directly lead to the probability distribution
\begin{align}
P^{\scriptscriptstyle\mathcal{E}}(x|\theta)=\operatorname{tr}(E_x(\theta)\rho E_x(\theta)^\dagger).\label{post}
\end{align}
If the post-selection isn't allowed, the state after the IO will become
 $\mathcal{E}_\theta(\rho)$. One can  operate a positive operator value measure (POVM) $\mathcal{M}=\{M_x\}$ on the state $\mathcal{E}_\theta(\rho)$, and obtain the probability distribution family as
\begin{align}
P_{\scriptscriptstyle\mathcal{M}}^{\scriptscriptstyle\mathcal{E}}(x|\theta)=\operatorname{tr}(M_x\mathcal{E}_\theta(\rho)),\label{gen}
\end{align}
where the subscript $\mathcal{M}$ denotes the general POVM.

The FI of distribution $P(x|\theta)$ is given by
\begin{align}
F(P,\theta_0)=\sum_xP(x|\theta_0)[\left.\frac{\partial \ln P(x|\theta)}{\partial\theta}\right|_{\theta_0}]^2,\label{Fisher}
\end{align}
and the QFI of $P_{\scriptscriptstyle\mathcal{M}}^{\scriptscriptstyle\mathcal{E}}(x|\theta)$ for any given $\theta_0$ can be written as
\begin{align}
F_{\scriptscriptstyle Q}(\rho,\mathcal{E},\theta_0)=\mathop{\max}\limits_{\mathcal{M}}F(P_{\scriptscriptstyle\mathcal{M}}^{\scriptscriptstyle\mathcal{E}},\theta_0).\label{QFI}
\end{align}
Based on above FI and QFI, we can establish two coherence measures, respectively, which will be given by the following two theorems.

\textbf{Theorem 1}.-The coherence of a state $\rho$ can be quantified by
the maximal FI for a given parameter $\theta_0$ as
 \begin{align}
C^{\scriptscriptstyle\theta_0}(\rho)=\mathop{\max}\limits_{\mathcal{E}\in G}
F(P^{\scriptscriptstyle\mathcal{E}},\theta_0),\label{d22}
\end{align}
where $F(P^{\scriptscriptstyle\mathcal{E}},\theta_0)$ is FI of distribution in Eq. (\ref{post}).

\textbf{Proof}:
We need to prove $C^{\scriptscriptstyle\theta_0}(\rho)$ satisfying $A1$-$A4$.

($A1$) \textit{Non-Negativity}.  If $\rho$ is incoherent, for any $\mathcal{E}$ and $x$, we have
\begin{align}
&E_x(\theta)\rho E_x(\theta)^\dagger\nonumber\\
=&\sum_nb_n^x(\theta)|g_x(n)\rangle\langle n|\rho\sum_mb_m^{x*}(\theta)|m\rangle\langle g_x(m)|\nonumber\\
=&\sum_{nm}b_n^x(\theta)b_m^{x*}(\theta)\rho_{nm}|g_x(n)\rangle\langle g_x(m)|\nonumber\\
=&\sum_n |b_n^x(\theta)|^2\rho_{nn}|g_x(n)\rangle\langle g_x(n)|,
\end{align}
which doesn't depend on $\theta$ due to $b_n^x(\theta)=c_n^xe^{ih_n^x(\theta)}$. Thus $P^{\scriptscriptstyle\mathcal{E}}(x|\theta)$
doesn't depend on $\theta$ either, which means \begin{align}
F(P^{\scriptscriptstyle\mathcal{E}},\theta_0)=\sum_{x}[\left.\frac{\partial P^{\scriptscriptstyle\mathcal{E}}(x|\theta)}{\partial\theta}\right|_{\theta_0}]^2\frac{1}{P^{\scriptscriptstyle\mathcal{E}}(x|\theta_0)}=0.\label{f1}
\end{align}
Eq. (\ref{f1})  leads to
$C^{\scriptscriptstyle\theta_0}(\rho)=0$.

Conversely, if a $d$-dimensional $\rho$ has non-zero off-diagonal entries, without loss of generality, one can let $\rho_{12}=|\rho_{12}|e^{i\alpha}$.
There exists an IO  $\{E_i\}\in G$,
\begin{align}
E_1(\theta)=&\frac{\sqrt{2}}{2}e^{i(\theta+\gamma)}|1\rangle\langle 1|+\frac{\sqrt{2}}{2}|1\rangle\langle 2|,\nonumber\\
E_2(\theta)=&-\frac{\sqrt{2}}{2}e^{i(\theta+\gamma)}|2\rangle\langle 1|+\frac{\sqrt{2}}{2}|2\rangle\langle 2|,\nonumber\\
E_3(\theta)=&\sum_{n=3}^d|n\rangle\langle n|,
\end{align}
with $\alpha+\theta_0+\gamma\in[-\pi/2,0)\bigcup(0,\pi/2]$, such that
\begin{align}
&P^{\scriptscriptstyle\mathcal{E}}(1|\theta_0)=\operatorname{tr}(E_1(\theta_0)\rho E_1(\theta_0)^\dagger)\neq0,\nonumber\\
&\partial_\theta \operatorname{tr}(E_1(\theta)\rho E_1(\theta)^\dagger )|_{\theta_0}\neq 0,
\end{align}
which obviously shows $C^{\scriptscriptstyle\theta_0}(\rho)\neq 0$ and $C^{\scriptscriptstyle\theta_0}(\rho)> 0$.

($A3$) \textit{Strong monotonicity.} Suppose $\rho$ undergoes an arbitrary IO
\begin{align}
K_l=&\sum_na_n^l|f_l(n)\rangle\langle n|,
\end{align}
the post-measurement ensemble $\{t_l,\rho_l\}$ reads
\begin{align}
t_l=\operatorname{tr}(K_l\rho K_l^\dagger), \rho_l=K_l\rho K_l^\dagger/t_l.
\end{align}
Let $\mathcal{E}^{(l)}=\{E_x^l(\theta)\}_x$ be the optimal IO for $\rho_l$ such that
\begin{align}
C^{\scriptscriptstyle\theta_0}(\rho_l)=F(P_l,\theta_0),
\end{align}
where $E^l_x(\theta)=\sum_nb_n^{lx}(\theta)|g_{lx}(n)\rangle\langle n|$ and
\begin{align}
P_l(x|\theta)
=&\operatorname{tr}(E_x^l(\theta)\rho_lE_x^{l}(\theta)^\dagger)\nonumber\\
=&\operatorname{tr}(E_x^l(\theta)K_l\rho K_l^\dagger E_x^{l}(\theta)^\dagger)/t_l\nonumber\\
=&P(x,l|\theta)/t_l.
\end{align}
Above $P(x,l|\theta)$ represents the probability distribution from
$\mathcal{E}'=\{E'_{xl}(\theta)\}_{xl}$ with
\begin{align}
&E'_{xl}(\theta)=E_x^l(\theta)K_l
=\sum_na_n^lb_{f_l(n)}^{lx}(\theta)|g_{lx}[f_l(n)]\rangle\langle n|,
\end{align}
which implies $\mathcal{E}'\in G$.
Therefore, one can arrive at
\begin{align}
&\sum_lt_lC^{\scriptscriptstyle\theta_0}(\rho_l)=\sum_{l}t_lF(P_l,\theta_0) \nonumber\\
=&\sum_lt_l\sum_{x\in S_l}[\left.\frac{\partial P_l(x|\theta)}{\partial\theta}\right|_{\theta_0}]^2\frac{1}{P_l(x|\theta_0)}\nonumber\\
=&\sum_lt_l\sum_{x\in S_l}[\left.\frac{\partial P(l,x|\theta)}{\partial\theta}\right|_{\theta_0}]^2\frac{1}{P(l,x|\theta_0)t_l}\nonumber\\
=&\sum_l\sum_{x\in S_l}[\left.\frac{\partial P(l,x|\theta)}{\partial\theta}\right|_{\theta_0}]^2\frac{1}{P(l,x|\theta_0)}\nonumber\\
=&F(P,\theta_0)\leq C^{\scriptscriptstyle\theta_0}(\rho),
\end{align}
where $S_l$ indicates the region of $x$ in $P_l$, and the last inequality is from that $\mathcal{E}'$ may not be the optimal one for $\rho$.

($A4$) \textit{Convexity.}
For any ensemble $\{t_i,\sigma_i\}$ with the corresponding mixed state
$
\rho=\sum_it_i\sigma_i
$,
let $\mathcal{E}=\{E_x(\theta)\}$ be the optimal IO for $\rho$ in the sense of
$
C^{\scriptscriptstyle\theta_0}(\rho)=F(P,\theta_0)
$
with
$
P(x|\theta)=
\operatorname{tr}(E_x(\theta)\rho E_x^{\dagger}(\theta))
$.
For the state $\sigma_i$, denote
\begin{align}
P_i(x|\theta)=
\operatorname{tr}(E_x(\theta)\sigma_i E_x^{\dagger}(\theta)),
\end{align}
then
\begin{align}
\sum_it_iP_i(x|\theta)
=&\sum_it_i\operatorname{tr}(E_x(\theta)\sigma_i E_x^{\dagger}(\theta))\nonumber\\
=&\operatorname{tr}(E_x(\theta)\rho E_x^{\dagger}(\theta))\nonumber\\
=&P(x|\theta).
\end{align}
However,
$\mathcal{E}$ may not be optimal for $\sigma_i$, which implies \begin{align}
C^{\scriptscriptstyle\theta_0}(\sigma_i)\geq F(P_i,\theta_0),
\end{align}
so one can immediately get
\begin{align}
\sum_it_iC^{\scriptscriptstyle\theta_0}(\sigma_i)&\geq \sum_it_iF(P_i,\theta_0)\nonumber\\
\geq&F(\sum_it_iP_i,\theta_0)=F(P,\theta_0)\nonumber\\
=&C^{\scriptscriptstyle\theta_0}(\rho),
\end{align}
where the second inequality is due to the convexity of FI.

Since ($A3$) and ($A4$) hold, it is natural that ($A2$) is satisfied. The proof is completed.
$\hfill\qedsymbol$

From Theorem 1, coherence could be quantified by FI of probability distribution in Eq. (\ref{post}), in some sense, this implies the connection between coherence and estimation accuracy for incoherent non-unitary parametrization. In fact, $G$ in the definition Eq. (\ref{d22}) could be replaced by its subset $G_1$ from the lemma below.

\textbf{Lemma 1}.-For any $\mathcal{E}=\{E_x(\theta)\}\in G$ , there always exists another $\mathcal{E}'=\{\tilde{E}_{x}(\theta)\}\in G_1$, such that
\begin{align}
F(P^{\scriptscriptstyle\mathcal{E}},\theta_0)\leq F(P^{\scriptscriptstyle\mathcal{E}'},\theta_0),
\end{align}
where $F(P^{\scriptscriptstyle\mathcal{E}},\theta_0)$ and
$F(P^{\scriptscriptstyle\mathcal{E}'},\theta_0)$ are FI of $P^{\scriptscriptstyle\mathcal{E}}(x|\theta)$ and $P^{\scriptscriptstyle\mathcal{E}'}(x|\theta)$ respectively.

\textbf{Proof:} Let  $\mathcal{E}=\{E_x(\theta)\}\in G$, one can rewrite $\{E_x(\theta)\}$ as
\begin{align}E_x(\theta)
=&\sum_nc_n^xe^{ih_n^x(\theta)}|g_x(n)\rangle\langle n|\nonumber\\
=&\sum_nc_n^x|g_x(n)\rangle\langle n|\sum_me^{ih_m^x(\theta)}|m\rangle\langle m|\nonumber\\
=&A_xU_x(\theta),
\end{align}
where $A_x=\sum_n$ $c_n^x|g_x(n)\rangle\langle n|$, and $U_x(\theta)=$ $\sum_m$ $e^{ih_m^x(\theta)}$ $|m\rangle\langle m|$. Thus we have\begin{align}
&E_{x}(\theta)^\dagger E_{x}(\theta)=U_{x}(\theta)^\dagger A_{x}^\dagger A_{x}U_{x}(\theta) \nonumber\\
&=U_{x}(\theta)^\dagger(\sum_i|\psi_i^{x}\rangle\langle\psi_i^{x}|)U_{x}(\theta) \nonumber\\
&=\sum_iU_{x}(\theta)^\dagger|\psi_i^{x}\rangle\langle\psi_i^{x}|U_{x}(\theta) \nonumber\\
&=\sum_i|\phi_i^{x}(\theta)\rangle\langle\phi_i^{x}(\theta)| =\sum_i\tilde{E}_{x,i}(\theta)^\dagger\tilde{E}_{x,i}(\theta),\label{77}
\end{align}
where $\sum_i|\psi_i^{x}\rangle\langle\psi_i^{x}|$ denotes the eigen-decomposition of $A_x^\dagger A_x$ (the eigenvalue is absorbed in $|\psi_i^{x}\rangle$), $|\phi_i^{x}(\theta)\rangle=U_{x}(\theta)^\dagger|\psi_i^{x}\rangle$ and
$
\tilde{E}_{x,i}(\theta)=|i\rangle\langle\phi_i^{x}(\theta)|
$. It is obvious that $\mathcal{E}'=
\{\tilde{E}_{x,i}(\theta)\}_{xi}\in G_1
$.
From Cauchy-Schwarz inequality ($|\langle v|w\rangle|^2\leq\langle v|v\rangle\langle w|w\rangle$)  \cite{steele2004cauchy}, one can obtain
\begin{align}
[\partial_{\theta}P(x|\theta)|_{\theta_0}]^2\leq\sum_i\frac{[\partial_\theta P_i(x|\theta)|_{\theta_0}]^2}{P_i(x|\theta_0)}\sum_iP_i(x|\theta_0),\label{pro}
\end{align}
where $P(x|\theta)=\operatorname{tr}(\rho E_{x}^\dagger E_{x}), P_i(x|\theta)=\operatorname{tr}(\rho \tilde {E}_{x,i}^\dagger \tilde {E}_{x,i})$,
thus
\begin{align}
\frac{[\partial_\theta P(x|\theta)|_{\theta_0}]^2}{P(x|\theta_0)}\leq
\sum_i\frac{[\partial_\theta P_i(x|\theta)|_{\theta_0}]^2}{P_i(x|\theta_0)},\label{resul}
\end{align}
the inequality holds for every $x$, which implies $F(P^{\scriptscriptstyle\mathcal{E}},\theta_0)\leq F(P^{\scriptscriptstyle\mathcal{E}'},\theta_0)$.
$\hfill\qedsymbol$

From the lemma, maximizing the FI over the set $G$ can be realized by the optimization over the set $G_1$, which effectively reduces the range of the optimized IO.

Theorem 1 mainly focuses on the FI with the related probability distribution
generated via the post-selective IO on a state.
Next, we would build another coherence measure defined by QFI with respect to parametrization in $G$,
\begin{align}
C_{\scriptscriptstyle Q}^{\scriptscriptstyle\theta_0}(\rho)=\mathop{\max}\limits_{\mathcal{E}\in G}F_{\scriptscriptstyle Q}(\rho,\mathcal{E},\theta_0).\label{CQFI}
\end{align}
To do this, we would first give a lemma.

\textbf{Lemma 2}.- The maximal QFI subject to parametrization in $G$ is upper bounded by the FI directly induced by the optimal post-selective IO parametrization process, namely,
\begin{align}
\mathop{\max}\limits_{\mathcal{E}\in
G}F_{\scriptscriptstyle Q}(\rho,\mathcal{E},\theta_0)\leq\mathop{\max}\limits_{\mathcal{E}\in G}
F(P^{\scriptscriptstyle\mathcal{E}},\theta_0),\label{QFI2}
\end{align}
where $P^{\scriptscriptstyle\mathcal{E}}$ is the distribution in Eq. (\ref{post}).

\textbf{Proof:}  Suppose $\tilde{\mathcal{E}}$ and $\mathcal{M}$ are the optimal parametrization and measurement for the optimal $F_{\scriptscriptstyle Q}$ respectively,
from Eq. (\ref{gen}), we have $P_{\scriptscriptstyle\mathcal{M}}^{\scriptscriptstyle\tilde{\mathcal{E}}}(x|\theta)
$ $=\operatorname{tr}(\sum_i|\psi_i^x\rangle\langle\psi_i^x|\tilde{\mathcal{E}}_\theta(\rho))$ $=\sum_iP_i(x|\theta)$ where $\sum_i|\psi_i^x\rangle\langle\psi_i^x|$
represents the eigen-decomposition of $M_x$. In particular, $P_i(x|\theta)=\left\langle\psi_i^x\right\vert\tilde{\mathcal{E}}_\theta(\rho)\left\vert\psi_i^x\right\rangle$, which can be rewritten as
\begin{align}
&P_i(x|\theta)=\operatorname{tr}(\left\vert i\right\rangle\left\langle\psi_i^x\right\vert\tilde{\mathcal{E}}_\theta(\rho)\left\vert\psi_i^x\right\rangle\left\langle i\right\vert)\nonumber\\
=&\sum_{ynn'}\operatorname{tr}(\left\vert i\right\rangle\left\langle\psi_i^x\right\vert b_n^y(\theta)|g_y(n)\rangle\langle n|\rho |n'\rangle\langle g_y(n')|b_{n'}^{y*}(\theta)\left\vert\psi_i^x\right\rangle\left\langle i\right\vert)\nonumber\\
=&\sum_{ynn'}\operatorname{tr}(b_{n}^{ixy}(\theta)\left\vert i\right\rangle\langle n|\rho |n'\rangle\left\langle i\right\vert b_{n'}^{ixy*}(\theta))\nonumber\\
=&\sum_y\operatorname{tr}(E_{ixy}(\theta)\rho E_{ixy}(\theta)^\dagger)=\sum_yP(ixy|\theta),
\end{align}
where $b_{n}^{ixy}(\theta)=\left\langle\psi_i^x\right\vert b_n^y(\theta)|g_y(n)\rangle$ and $E_{ixy}(\theta)=\sum_{n}b_{n}^{ixy}(\theta)\left\vert i\right\rangle\langle n|$.
It is obvious that $\mathcal{E}'_\theta=\{E_{ixy}(\theta)\}\in G$, then
\begin{align}
&\mathop{\max}\limits_{\mathcal{E}}F_{\scriptscriptstyle Q}(\rho,\mathcal{E},\theta_0)=F(P_{\scriptscriptstyle\mathcal{M}}^{\scriptscriptstyle\tilde{\mathcal{E}}},\theta_0)\nonumber\\
&=\sum_x\frac{[\partial_\theta P_{\scriptscriptstyle\mathcal{M}}^{\scriptscriptstyle\tilde{\mathcal{E}}}(x|\theta)|_{\theta_0}]^2}{P_{\scriptscriptstyle\mathcal{M}}^{\scriptscriptstyle\tilde{\mathcal{E}}}(x|\theta_0)}=
\sum_x\frac{[\sum_i\partial_\theta P_i(x|\theta)|_{\theta_0}]^2}{\sum_iP_i(x|\theta_0)}\nonumber\\
&\leq\sum_{ix}\frac{[\partial_\theta P_i(x|\theta)|_{\theta_0}]^2}{P_i(x|\theta_0)}
=\sum_{ix}\frac{[\sum_y\partial_\theta P(ixy|\theta)|_{\theta_0}]^2}{\sum_yP(ixy|\theta_0)}\nonumber\\
&\leq\sum_{ixy}\frac{[\partial_\theta P(ixy|\theta)|_{\theta_0}]^2}{P(ixy|\theta_0)}=F(P,\theta_0)\leq\mathop{\max}\limits_{\mathcal{E}}
F(P^{\scriptscriptstyle\mathcal{E}},\theta_0),\label{proo}
\end{align}
where $P$ is distribution from $\mathcal{E}'_\theta$
, the first two inequality could be derived based on Cauchy-Schwarz inequality, and the derivation process is similar as Eq. (\ref{pro}) and (\ref{resul}), namely,
from
\begin{align*}
[\sum_i\partial_\theta P_i(x|\theta)|_{\theta_0}]^2\leq \sum_i\frac{[\partial_\theta P_i(x|\theta)|_{\theta_0}]^2}{P_i(x|\theta_0)}\sum_iP_i(x|\theta_0)
\end{align*}
we could obtain the first inequality,
and from
\begin{align*}
[\sum_y\partial_\theta P(ixy|\theta)|_{\theta_0}]^2\leq\sum_y\frac{[\partial_\theta P(ixy|\theta)|_{\theta_0}]^2}{P(ixy|\theta_0)}\sum_yP(ixy|\theta_0)
\end{align*}
we could reach the second inequality.
Thus one can complete the proof.$\hfill\qedsymbol$

Next, we show that
$C_{\scriptscriptstyle Q}^{\scriptscriptstyle\theta_0}(\rho)$ in Eq. (\ref{CQFI}) is equivalent to $C^{\scriptscriptstyle\theta_0}(\rho)$, and can also quantify the quantum coherence of $\rho$.

\textbf{Theorem 2}.-For a given density matrix $\rho$,
\begin{equation}
C_{\scriptscriptstyle Q}^{\scriptscriptstyle\theta_0}(\rho)=C^{\scriptscriptstyle\theta_0}(\rho).\label{dengshi}
\end{equation}

\textbf{Proof:}
From Lemma 1, $C^{\scriptscriptstyle\theta_0}(\rho)$ could be written as
\begin{align}
C^{\scriptscriptstyle\theta_0}(\rho)=\mathop{\max}\limits_{\mathcal{E}\in G_1}
F(P^{\scriptscriptstyle\mathcal{E}},\theta_0).
\end{align}
Suppose $\mathcal{E}=\{E_z(\theta)\}$ is the optimal operation in $G_1$, such that
\begin{align}
C^{\scriptscriptstyle\theta_0}(\rho)=F(P^{\scriptscriptstyle\mathcal{E}},\theta_0),
\end{align}
here
\begin{align}
&P^{\scriptscriptstyle\mathcal{E}}(z|\theta)=\operatorname{tr}(E_z(\theta)\rho E_z(\theta)^\dagger).\label{pl2}
\end{align}
and $Rank[E_z(\theta)^\dagger E_z(\theta)]=1$,
without lose of generality, $E_z(\theta)$ could be written as
\begin{align}
E_z(\theta)=|z\rangle\langle\phi_z(\theta)|.
\end{align}
Denote
\begin{align}
P^{\scriptscriptstyle\mathcal{E}}_{\scriptscriptstyle\mathcal{P}}(z|\theta)=\operatorname{tr}(|z\rangle\langle z|\mathcal{E}_\theta(\rho)|z\rangle\langle z|),
\end{align}
where $\mathcal{P}$ indicates the projective measurements on the parameterized state.
Note that
\begin{align}
&P^{\scriptscriptstyle\mathcal{E}}(z|\theta)=\operatorname{tr}(E_z\rho E_z^\dagger)\nonumber\\
=&\operatorname{tr}(|z\rangle\langle z|(\sum_{z'}E_{z'}\rho E_{z'}^\dagger)|z\rangle\langle z|) \nonumber\\
=&\operatorname{tr}(|z\rangle\langle z|\mathcal{E}_\theta(\rho)|z\rangle\langle z|)=P^{\scriptscriptstyle\mathcal{E}}_{\scriptscriptstyle\mathcal{P}}(z|\theta),
\end{align}
thus
\begin{align}
&C^{\scriptscriptstyle\theta_0}(\rho)=F(P^{\scriptscriptstyle\mathcal{E}},\theta_0)=F(P^{\scriptscriptstyle\mathcal{E}}_{\scriptscriptstyle\mathcal{P}},\theta_0)
\leq \mathop{\max}\limits_{\mathcal{M}}F(P_{\scriptscriptstyle\mathcal{M}}^{\scriptscriptstyle\mathcal{E}},\theta_0) \nonumber\\
=&F_{\scriptscriptstyle Q}(\rho,\mathcal{E},\theta_0)\leq\mathop{\max}\limits_{\mathcal{E}\in G}F_{\scriptscriptstyle Q}(\rho,\mathcal{E},\theta_0)=C_{\scriptscriptstyle Q}^{\scriptscriptstyle\theta_0}(\rho).\label{project}
\end{align}

Conversely, from Lemma 2, we can immediately reach that
\begin{align}
C^{\scriptscriptstyle\theta_0}(\rho)\geq C_{\scriptscriptstyle Q}^{\scriptscriptstyle\theta_0}(\rho),
\end{align}
thus one can get the $C_{\scriptscriptstyle Q}^{\scriptscriptstyle\theta_0}(\rho)=C^{\scriptscriptstyle\theta_0}(\rho)$, which finishes the proof. $\hfill\qedsymbol$

We have shown that the coherence measures based on QFI and FI subject to the post-selective parametrization are equivalent to each other.
The most distinct advantage of this type of coherence measure is that it can be straightforwardly connected with the parameter estimation process in terms of the Cram\'{e}r-Rao bound \cite{rao1945information,PhysRevLett.72.3439,hayashi2006quantum}.

Let's consider an incoherent non-unitary parametrization $\mathcal{E}=\{E_x(\theta)\}\in G$ on $\rho$ as introduced previously,  then one will obtain a probability distribution $P^{\scriptscriptstyle\mathcal{E}}_{\scriptscriptstyle\mathcal{M}}(x|\theta)$ through a POVM on $\rho_\theta$ or obtain $P^{\scriptscriptstyle\mathcal{E}}(x|\theta)$ directly through post-selection of $\mathcal{E}$.
With maximum likelihood estimators $\hat{\theta}_{\scriptscriptstyle\mathcal{M}}$ with respect to $P^{\scriptscriptstyle\mathcal{E}}_{\scriptscriptstyle\mathcal{M}}$ or $\hat{\theta}$ with respect to $P^{\scriptscriptstyle\mathcal{E}}$, the Cram\'{e}r-Rao bound can be asymptotically attained.
That is, the mean square error
$(\delta\hat{\theta}_{\scriptscriptstyle\mathcal{M}})^2=E[(\hat{\theta}_{\scriptscriptstyle\mathcal{M}}-\theta)^2]$ and $(\delta\hat{\theta})^2=E[(\hat{\theta}-\theta)^2]$ approach $\frac{1}{nF}$ in the asymptotic sense, where $E$ indicates the expectation value, $\theta$ is the true value and $n$ denotes the runs of detection.
Thus in the asymptotic limit, the estimation accuracy $\frac{1}{n(\delta\hat{\theta}_{\scriptscriptstyle\mathcal{M}})^2}$ approaches $F(P^{\scriptscriptstyle\mathcal{E}}_{\scriptscriptstyle\mathcal{M}},\theta)$, which is naturally bounded by $C_{\scriptscriptstyle \mathcal{Q}}^{\scriptscriptstyle\theta}(\rho)$ based on Eq. (26). In particular, the bound $C_{\scriptscriptstyle \mathcal{Q}}^{\scriptscriptstyle\theta}(\rho)$  can be asymptotically achieved with the optimal parametrization process and optimal POVM. Similarly,  $\frac{1}{n(\delta\hat{\theta})^2}$ approaches $F(P^{\scriptscriptstyle\mathcal{E}},\theta)$ in the asymptotic scenario, and simultaneously reach $ C^{\scriptscriptstyle\theta}(\rho)$ in an asymptotic sense with an optimal parametrization process. Note that the two measures are equivalent, therefore our coherence measure can be understood as the optimal accuracy  through two different estimation processes as well as the corresponding incoherent non-unitary parametrization.

In fact, the optimized $\mathcal{M}$ in $C_{\scriptscriptstyle Q}^{\scriptscriptstyle\theta_0}$ (Eq. (\ref{CQFI})) can be replaced
by $\mathcal{P}$, the projective measurement on the preferred basis. In this sense, the above two coherence measures have an equivalent expression as $C_{\scriptscriptstyle\mathcal{P}}^{\scriptscriptstyle\theta_0}(\rho)=\mathop{\max}\limits_{\mathcal{E}\in G}
F(P^{\scriptscriptstyle\mathcal{E}}_{\scriptscriptstyle\mathcal{P}},\theta_0)$.
This can be understood as follows.
We first have $C^{\scriptscriptstyle\theta_0}(\rho)\leq C_{\scriptscriptstyle\mathcal{P}}^{\scriptscriptstyle\theta_0}(\rho)$ from the second equality in Eq. (\ref{project}). Note that $\mathcal{M}$ in $C_{\scriptscriptstyle Q}^{\scriptscriptstyle\theta_0}(\rho)$ contains projective measurement, which implies $C_{\scriptscriptstyle Q}^{\scriptscriptstyle\theta_0}(\rho)\geq C_{\scriptscriptstyle\mathcal{P}}^{\scriptscriptstyle\theta_0}(\rho)$. Combine the above two inequalities with Theorem 2, one can reach $C_{\scriptscriptstyle\mathcal{P}}^{\scriptscriptstyle\theta_0}(\rho)
=C^{\scriptscriptstyle\theta_0}(\rho)=C_{\scriptscriptstyle Q}^{\scriptscriptstyle\theta_0}(\rho)$.
Although they are identical in value, they imply different details of operational meanings and give us different ways to understand coherence.

\section{Connection with QFI based on unitary parametrization}

Although the coherence measure has obvious operation meaning based on quantum metrology, an analytically computable expression seems not to be easy. Next, we will show that for a 2-dimensional quantum state, the analytic result could be obtained, and the coherence measure can be realized by FI with unitary parametrization. However,
our measure is not equivalent to that based on unitary parametrization in high-dimensional cases, which is proved later.

\textbf{Theorem 3}.-For a 2-dimensional state $\rho$,  the coherence based on Theorem 1 can be given as
\begin{align}
C^{\scriptscriptstyle \theta_0}(\rho)=F_{\scriptscriptstyle Q}(\rho,U_\theta,\theta_0),
\end{align}
where $F_{\scriptscriptstyle Q}$ is QFI of $\rho$ subject to unitary parametrization $U_\theta=e^{i\theta}|1\rangle\langle 1|+|2\rangle\langle 2|$.

\textbf{Proof:} For qubit states $\rho$, let the IO $\{E_x\}\in G$ read
\begin{align}
E_x(\theta)=a_1'^xe^{i h_1'^x(\theta)}|f_x(1)\rangle\langle 1|+a_2'^xe^{ih_2'^x(\theta)}|f_x(2)\rangle\langle 2|, \label{123}
\end{align}
where $a_1'^x$ or $a_2'^x$ may be zero. The Kraus operator could be written as
\begin{align}
E_x(\theta)=a_1^xe^{ih_1^x(\theta)}|f_x(1)\rangle\langle 1|+a_2^xe^{i h_2^x(\theta)}|f_x(2)\rangle\langle 2|,
\end{align}
where $a_j^x=a_j'^xe^{ih_j'^x(\theta_0)}$ and $h_j^x(\theta)=h_j'^x(\theta)-h_j'^x(\theta_0)$ for $j=1,2$.
According to Lemma 1 and its proof,
the optimal IO can be rank-1 with the form $\{\left\vert i\right\rangle\left\langle \psi_i^x(\theta)\right\vert\}$, which means $f_x(1)=f_x(2)$ for any $x$. Then we have
\begin{align}
&P(x|\theta)\nonumber\\
=&\operatorname{tr}
(|a_1^x|^2\rho_{11}|f_x(1)\rangle\langle f_x(1)|+|a_2^x|^2\rho_{22}|f_x(2)\rangle\langle f_x(2)| \nonumber\\
+&\rho_{12}a_1^xa_2^{x*}e^{i[h_1^x(\theta)-h_2^x(\theta)]}|f_x(1)\rangle\langle f_x(2)| \nonumber\\
+&\rho_{21}a_1^{x*}a_2^{x}e^{-i[h_1^x(\theta)-h_2^x(\theta)]}|f_x(2)\rangle\langle f_x(1)|) \nonumber\\
=&|a_1^x|^2\rho_{11}+|a_2^x|^2\rho_{22}+\rho_{12}a_1^xa_2^{x*}e^{i[h_1^x(\theta)-h_2^x(\theta)]} \nonumber\\
+&\rho_{21}a_1^{x*}a_2^{x}e^{-i[h_1^x(\theta)-h_2^x(\theta)]},
\end{align}
thus
\begin{align}
&F(P,\theta_0)=
\sum_x\frac{[2Im(\rho_{12}a_1^xa_2^{x*})]^2[\partial_\theta h_1^x(\theta)|_{\theta_0}-\partial_\theta h_2^x(\theta)|_{\theta_0}]^2}{|a_1^x|^2\rho_{11}+|a_2^x|^2\rho_{22}
+2Re(\rho_{12}a_1^xa_2^{x*})} \nonumber\\
\leq&\sum_x\frac{[2Im(\rho_{12}a_1^xa_2^{x*})]^2}{|a_1^x|^2\rho_{11}+|a_2^x|^2\rho_{22}+2Re(\rho_{12}a_1^xa_2^{x*})},
\end{align}
where the inequality could be saturated by the function taken as $h_1^x(\theta)=\theta$, $h_2^x(\theta)=0$,  and the corresponding IO reads
$E_x(\theta)=K_xU_\theta$
with
\begin{align}
K_x=&a_1^x|f_x(1)\rangle\langle 1|+a_2^x|f_x(2)\rangle\langle 2|,\nonumber\\
U_\theta=&e^{i\theta}|1\rangle\langle 1|+|2\rangle\langle 2|,
\end{align}
where $f_x(1)=f_x(2)$ and $\{K_x\}\in G_1$. In this sense, the probability distribution can be rewritten as
\begin{align}
&P^{\scriptscriptstyle\mathcal{E}}(x|\theta)=\operatorname{tr}(E_x(\theta)\rho E_x(\theta)^\dagger)=\operatorname{tr}(K_xU_\theta\rho U_\theta^\dagger K_x^\dagger)\nonumber\\
=&\operatorname{tr}(U_\theta\rho U_\theta^\dagger K_x^\dagger K_x)=P_{\scriptscriptstyle\mathcal{M}}(x|\theta).\label{imp}
\end{align}
above $P_{\scriptscriptstyle\mathcal{M}}$ can be understood as distribution generated by a unitary parametrization $U_\theta$ followed by a rank-1 POVM $\mathcal{M}=\{K_x^\dagger K_x\}$.
Considering the above optimal IO, one can arrive at
\begin{align}
C^{\theta_0}(\rho)&=\mathop{\max}\limits_{\mathcal{E}\in G_1}
F(P^{\scriptscriptstyle\mathcal{E}},\theta_0)\nonumber\\
&=\mathop{\max}\limits_{\mathcal{M}}F(P_{\scriptscriptstyle\mathcal{M}},\theta_0)=F_{\scriptscriptstyle Q}(\rho,U_\theta,\theta_0),\label{T4}
\end{align}
which finishes the proof.$\hfill\qedsymbol$

In fact, in general high-dimensional case, $C^{\scriptscriptstyle\theta_0}$ is distinct from FI with unitary parametrization.
To demonstrate the difference, we will give a concrete example.
Consider a state with maximal coherence,
\begin{align}
|\phi\rangle=(\frac{1}{\sqrt{3}},\frac{1}{\sqrt{3}},\frac{1}{\sqrt{3}})^T,\label{MCS}
\end{align}
and the parametrization $\mathcal{E}=\{E_x(\theta)\}$ expressed as
\begin{align}
E_x(\theta)=&a_1^xe^{ih_1^x\theta}|f_x(1)\rangle\langle 1|
+a_2^xe^{ih_2^x\theta}|f_x(1)\rangle\langle 2|\nonumber\\
+&a_3^xe^{ih_3^x\theta}|f_x(1)\rangle\langle 3|,
\end{align}
with  $a_n^x$ and $h_n^x$ ($x=1,\cdots,9$) to be given at the end.
Denote $\rho=|\phi\rangle\langle\phi|$. The probability distribution is
\begin{align}
P(x|0)=&\operatorname{tr}(E_x(\theta)|\phi\rangle\langle\phi| E_x(\theta)^\dagger)\nonumber\\
=&\rho_{11}|a_1^x|^2+\rho_{22}|a_2^x|^2+\rho_{33}|a_3^x|^2\nonumber\\
+&2\operatorname{Re}[\rho_{12}a_1^xa_2^{x*}+\rho_{12}a_2^xa_3^{x*}+\rho_{31}a_3^xa_1^{x*}],
\end{align}
and
\begin{align}
\partial_\theta P(x|\theta)|_0=&2\operatorname{Im}[\rho_{12}a_1^xa_2^{x*}(h_1^x-h_2^x)
+\rho_{23}a_2^xa_3^{x*}(h_2^x-h_3^x)\nonumber\\
+&\rho_{31}a_3^xa_1^{x*}(h_3^x-h_1^x)].
\end{align}
Therefore, the corresponding FI reads
\begin{align}
F(P^\mathcal{E},0)=\sum_x\frac{[\partial_\theta P(x|\theta)|_0]^2}{P(x|0)}=0.9410.
\end{align}
From the definition, we have $C^0(\rho)\geq F(P^\mathcal{E},0)$.

To compare our measure with QFI subject to the optimal unitary parametrization in $G$, we calculate $\mathop{\max}\limits_{U_\theta\in G}F_{\scriptscriptstyle Q}(|\phi\rangle,U_\theta,0)$, where $U_\theta$ is the unitary operator expressed as
\begin{align}
U_\theta=\sum_ne^{ih_n(\theta)}|n\rangle\langle n|
\end{align}
with $\partial_\theta h_n(\theta)\in[0,1]$ (based on Appendix A, other cases with different range of $\partial_\theta h_n(\theta)$ lead the same conclusion).
When eigenvalues of the parameterized state $U_\theta\rho U_\theta^\dagger$ are parameter-independent, QFI could be calculated from the following equation  \cite{PhysRevA.88.043832,PhysRevA.87.022337},
\begin{align}
F_{\scriptscriptstyle Q}(\rho,U_\theta,\theta_0)
=\sum_{ij}\frac{2(P_i-P_j)^2}{P_i+P_j}|\langle\varphi_i|\partial_\theta\varphi_j\rangle|^2,
\end{align}
where $\{P_i\}$ and $\{|\varphi_i\rangle\}$ denote the eigenvalues and eigenvectors of $U_\theta\rho U_\theta^\dagger$ respectively, and we use $|\partial_\theta\varphi_j\rangle$ to briefly express the partial derivative $\frac{\partial|\varphi_j\rangle}{\partial\theta}|_{\theta_0}$. Besides, the terms with $P_i=P_j=0$ are not included in the summation.
In addition, for a pure state $\rho=|\psi\rangle\langle\psi|$, let $\{|\psi_i\rangle\}$ be the basis vectors satisfying $|\psi\rangle=|\psi_1\rangle$, then the corresponding $P_1=1$ and residual eigenvalues $P_i$ ($i\neq1$) are zero. Then the eigenvectors of $U_\theta\rho U_\theta^\dagger$ are $\{U_\theta|\psi_i\rangle\}$. Denote
\begin{align}
H_\theta=\sum_n\partial_\theta h_n(\theta)|n\rangle\langle n|,
\end{align}
we have
\begin{align}
&F_{\scriptscriptstyle Q}(|\psi\rangle,U_\theta,\theta_0)\nonumber\\=&\sum_{i}\frac{2(1-P_i)^2}{1+P_i}
\langle\psi|U_{\theta_0}^\dagger U_{\theta_0} H_{\theta_0}|\psi_i\rangle\langle\psi_i|H_{\theta_0} U_{\theta_0}^\dagger U_{\theta_0}|\psi\rangle\nonumber\\
+&\sum_{i}\frac{2(P_i-1)^2}{P_i+1}
\langle\psi_i|U_{\theta_0}^\dagger U_{\theta_0} H_{\theta_0}|\psi\rangle\langle\psi|H_{\theta_0} U_{\theta_0}^\dagger U_{\theta_0}|\psi_i\rangle\nonumber\\
=&4\langle\psi|H_{\theta_0}\sum_{i}|\psi_i\rangle\langle\psi_i|H_{\theta_0}|\psi\rangle
-4\langle\psi|H_{\theta_0}|\psi\rangle\langle\psi|H_{\theta_0}|\psi\rangle\nonumber\\
=&4\langle\psi|H_{\theta_0}^2|\psi\rangle-4\langle\psi|H_{\theta_0}|\psi\rangle^2.
\end{align}
This result does not depend on the choice of $|\psi_i\rangle$ as long as $|\psi\rangle=|\psi_1\rangle$, and the optimal QFI $\mathop{\max}\limits_{U_\theta\in G}F_{\scriptscriptstyle Q}(|\phi\rangle,U_\theta,0)$ can be calculated as
\begin{align}
\mathop{\max}\limits_{U_\theta\in G}F_{\scriptscriptstyle Q}(|\phi\rangle,U_\theta,0)=&\mathop{\max}\limits_{H\in S}4\langle\phi|H^2|\phi\rangle-4\langle\phi|H|\phi\rangle^2\nonumber\\
=&8/9,
\end{align}
where $|\phi\rangle$ is the $3-$dimensional MCS in Eq. (\ref{MCS}), $S$ is the set of operator $H=h_1|1\rangle\langle1|+h_2|2\rangle\langle2|+h_3|3\rangle\langle3|$ ($h_i\in[0,1]$).
Thus $C^0(\rho)>\mathop{\max}\limits_{U_\theta\in G}F(|\phi\rangle,U_\theta,0)$, which indicates that $C^{\scriptscriptstyle\theta_0}$ is different from FI with unitary parametrization.

Finally, we'd like to present all the coefficients of $E_x$ in above calculation by defining
$
A^x=[a_1^x,a_2^x,a_3^x],
$
where
\begin{align*}
A^1&=[0,\sqrt{0.4},\sqrt{0.6}]/\sqrt{3},
\end{align*}
\begin{align*}
A^2&=[0,\sqrt{0.4}e^{-i2\pi/3},\sqrt{0.6}e^{i2\pi/3}]/\sqrt{3},\nonumber\\
\end{align*}
\begin{align*}
A^3&=[0,\sqrt{0.4}e^{-i4\pi/3},\sqrt{0.6}e^{i4\pi/3}]/\sqrt{3},
\end{align*}
\begin{align*}
A^4&=[\sqrt{0.4},\sqrt{0.6},           0]/\sqrt{3},
\end{align*}
\begin{align*}
A^5&=[\sqrt{0.4},\sqrt{0.6}e^{i2\pi/3},0]/\sqrt{3},
\end{align*}
\begin{align*}
A^6&=[\sqrt{0.4},\sqrt{0.6}e^{i4\pi/3},0]/\sqrt{3},
\end{align*}
\begin{align*}
A^7&=[\sqrt{0.6}, 0, \sqrt{0.4}]/\sqrt{3},
\end{align*}
\begin{align*}
A^8&=[\sqrt{0.6}, 0, \sqrt{0.4}e^{i2\pi/3}]/\sqrt{3},
\end{align*}
\begin{align}
A^9&=[\sqrt{0.6}, 0, \sqrt{0.4}e^{i4\pi/3}]/\sqrt{3}.
\end{align}
In addition,
\begin{align*}
h_1^x=0,\ h_2^x=1,\ h_3^x=0,\  x=1,2,3
\end{align*}
\begin{align}
h_1^x=1,\ h_2^x=0,\ h_3^x=0,\  x=4,\cdots,9.
\end{align}

\section{Conclusions}
In this paper, we have established coherence measures based on FI subject to the incoherent non-unitary parametrization process. The coherence measure could be
defined by two forms based on FI  or QFI, which both imply the direct operational meaning by the connection with the parameter estimation accuracy. In addition, we compare our measure with QFI in unitary parametrization and find that in the qubit case, our coherence measure can be equivalently understood through unitary parametrization, and can be analytically calculated.
Our coherence also
sheds new light on the roles of the non-unitary parametrization process.

\section*{Acknowledgements}
This work was supported by Scientific Research Foundation for the PhD (Zhejiang Ocean University, No. 11065091222), the National Natural Science
Foundation of China under Grant No.12175029, No.11775040, and No. 12011530014.

\appendix

\section{Region of $\partial_\theta h(\theta)$}

In the previous sections, $C^{\theta_0}_{\scriptscriptstyle Q}$ and $C^{\theta_0}$ are defined under a certain condition $\frac{\partial h_n^x(\theta)}{\partial\theta}\in[0,1]$.  In fact, measures defined under other conditions can be transformed to the original $C^{\theta_0}$.

We first consider the case that $\frac{\partial h_n^x(\theta)}{\partial\theta}$ is finite, and suppose $\mathop{\max}\limits_{n,x}|\frac{\partial h_n^x(\theta)}{\partial\theta}|\leq k$ ($k$ is finite). Denote $\tilde{C}_k^{\theta_0}$ as the function defined in a similar way as $C^{\theta_0}$ (in Theorem 1) but with the different condition $\mathop{\max}\limits_{n,x}|\frac{\partial h_n^x(\theta)}{\partial\theta}|\leq k$, and $G^{(k)}$ as the set of the corresponding channels, namely,
\begin{align}
\tilde{C}_k^{\theta_0}(\rho)=\mathop{\max}\limits_{\mathcal{E}\in G^{(k)}}F(\tilde{P}^\mathcal{E},\theta_0),
\end{align}
where
\begin{align}
\tilde{P}^\mathcal{E}(x|\theta)=\operatorname{tr}(\tilde{E}_x(\theta)\rho\tilde{E}_x(\theta)^\dagger),\nonumber\\
\tilde{E}_x(\theta)=\sum_na_n^xe^{ih_n^x(\theta)}|f_x(n)\rangle\langle n|,\label{r3}
\end{align}
and $\mathop{\max}\limits_{n,x}|\frac{\partial h_n^x(\theta)}{\partial\theta}|\leq k$.\\
We find that $\tilde{C}_k^{\theta_0}$ has a connection with the previous coherence measure.

\textbf{Lemma 3}.-The function $\tilde{C}_k^{\theta_0}$ satisfies that
\begin{align}
C^{\gamma_0}(\rho)=\frac{1}{4k^2}\tilde{C}_k^{\theta_0}(\rho),\label{r2}
\end{align}
where $\gamma_0=2k\theta_0$.

In this sense, investigation under condition $\frac{\partial h_n^x(\theta)}{\partial\theta}\in[0,1]$ could cover all other situations where $k$ is finite.
Next, we give a brief proof.

\textbf{Proof:}
Suppose $\{E_x\}$ are Kraus operators of channel in $G^{(\frac{1}{2})}$, namely,
\begin{align}
E_x(\gamma)=\sum_na_n^xe^{iu_n^x(\gamma)}|f_x(n)\rangle\langle n|,
\end{align}
where $|\partial_\gamma u_n^x|\leq\frac{1}{2}$. Denote
\begin{align}
\hat{E}_x(\gamma)=e^{i\gamma/2}E_x(\gamma)=\sum_na_n^xe^{iv_n^x(\gamma)}|f_x(n)\rangle\langle n|,
\end{align}
where $v_n^x(\gamma)=u_n^x(\gamma)+\gamma/2$, thus $\partial_\gamma v_n^x\in[0,1]$. The two channels lead to identical effects, that is
\begin{align}
\hat{E}_x(\gamma)\rho \hat{E}_x(\gamma)^\dagger
=&e^{i\gamma/2}E_x(\gamma)\rho e^{-i\gamma/2}E_x(\gamma)^\dagger\nonumber\\
=&E_x(\gamma)\rho E_x(\gamma)^\dagger,
\end{align}
from this, we have
\begin{align}
C^{\gamma_0}(\rho)=\tilde{C}_{\scriptscriptstyle 1/2}^{\gamma_0}(\rho).\label{eq1}
\end{align}
Consider $\tilde{E}_x$ in Eq. (\ref{r3}),
denote $S_l^x$ as the set satisfying $f_x(n)=l$ when $n\in S_l^x$, then
\begin{align}
\tilde{P}(x|\theta)=\sum_l\sum_{n,n'\in S_l^x}\rho_{nn'}a_n^xa_{n'}^{x*}e^{i[h_n^x(\theta)-h_{n'}^x(\theta)]},
\end{align}
and
\begin{align}
\partial_{\theta}\tilde{P}(x|\theta)|_{\theta_0}=&\sum_l\sum_{n,n'\in S_l^x}\{\rho_{nn'}a_n^xa_{n'}^{x*}e^{i[h_n^x(\theta_0)-h_{n'}^x(\theta_0)]}\nonumber\\
&\times i
[\partial_\theta h_n^x(\theta)|_{\theta_0}-\partial_\theta h_{n'}^x(\theta)|_{\theta_0}]\}.
\end{align}
Denote $\gamma=2k\theta$, then
\begin{align}
\tilde{P}(x|\theta)=\sum_l\sum_{n,n'\in S_l^x}\rho_{nn'}a_n^xa_{n'}^{x*}e^{i[h_n^x(\frac{\gamma}{2k})-h_{n'}^x(\frac{\gamma}{2k})]},
\end{align}
thus $\tilde{P}(x|\theta)$ could be rewritten as $P(x|\gamma)$.
Define $g_n^x(\gamma)=h_n^x(\frac{\gamma}{2k})$, then  $\partial_\gamma g_n^x(\gamma)=\frac{\partial_\theta h_n^x(\theta)}{2k}$, thus $|\partial_\gamma g_n^x(\gamma)|\leq\frac{1}{2}$. In addition,
\begin{align}
&\partial_{\gamma}P(x|\gamma)|_{\gamma_0}\nonumber\\
=&\sum_l\sum_{n,n'\in S_l^x}\{\rho_{nn'}a_n^xa_{n'}^{x*}e^{i[h_n^x(\frac{\gamma_0}{2k})-h_{n'}^x(\frac{\gamma_0}{2k})]}\nonumber\\
\times &i[\partial_\gamma h_n^x(\frac{\gamma}{2k})|_{\gamma_0}-\partial_\gamma h_{n'}^x(\frac{\gamma}{2k})|_{\gamma_0}]\}\nonumber
\end{align}

\begin{align}
=&\frac{1}{2k}\sum_l\sum_{n,n'\in S_l^x}\{\rho_{nn'}a_n^xa_{n'}^{x*}e^{i[h_n^x(\theta_0)-h_{n'}^x(\theta_0)]}\nonumber\\
\times&i[\partial_\theta h_n^x(\theta)|_{\theta_0}-\partial_\theta h_{n'}^x(\theta)|_{\theta_0}]\}\nonumber\\
=&\frac{1}{2k}\partial_\theta\tilde{P}(x|\theta)|_{\theta_0},
\end{align}
where $\gamma_0=2k\theta_0$.
Thus
\begin{align}
&F(P,\gamma_0)=\sum_x\frac{[\partial_\gamma P(x|\gamma)|_{\gamma_0}]^2}{P(x|\gamma_0)}\nonumber\\
=&\sum_x\frac{[\partial_\theta \tilde{P}(x|\theta)|_{\theta_0}]^2}{4k^2\tilde{P}(x|\theta_0)}
=\frac{F(\tilde{P},\theta_0)}{4k^2},
\end{align}
combine it with Eq. (\ref{eq1}), we have
\begin{align}
C^{\gamma_0}(\rho)=\frac{1}{4k^2}\tilde{C}^{\theta_0}(\rho).
\end{align}
$\hfill\qedsymbol$

Above proof shows that if $\partial_\theta h_n^x(\theta)$ is finite, the investigation under condition $\partial_\theta h_n^x(\theta)\in[0,1]$ could cover all other general cases. However, if $\partial_\theta h_n^x(\theta)$ is infinite, FI and $C^{\scriptscriptstyle\theta_0}(\rho)$ will be infinite. Besides, physical models generally lead to a finite $\partial_\theta h_n^x(\theta)$, for example, the parametrization Ramsey interferometer could be written as $U_\theta=\operatorname{exp}(-i\theta J_z)$, the corresponding $\partial_\theta h_n^x(\theta)$ is finite ($J_z$ is the $z-$component of the total angular momentum). Thus we could focus on finite case.

\section{Dilation of the Optimal Channel in $G$}
For the estimation process in Eq. (\ref{post}), the optimal channel in $G_1$ could be written as
\begin{align}
E_x(\theta)=\sum_nb_n^x(\theta)|1\rangle\langle n|,
\end{align}
denote $\mathcal{H}_A$ ($d$-dimension) as the space for it.
Assume $|x_{\scriptscriptstyle B}\rangle$ are basis vectors in another space $\mathcal{H}_B$ ($L$-dimension), and we construct the following states in $\mathcal{H}_B$,
\begin{align}
|\psi_{\scriptscriptstyle B}^n\rangle=\sum_{x=1}^Lb_n^x(\theta)|x_{\scriptscriptstyle B}\rangle,\ \ n=1,2,\dots,d.\label{vec}
\end{align}
From $\sum_xE_x^\dagger E_x=\mathcal{I}$, we have $\langle\psi_{\scriptscriptstyle B}^m|\psi_{\scriptscriptstyle B}^n\rangle=\delta_{nm}$. With states in Eq. (\ref{vec}), one can always find the other $|\psi_{\scriptscriptstyle B}^x\rangle$ ($x=d+1,\dots,L$) to form a set of basis together in $\mathcal{H}_B$.
Denote
\begin{align}
U^B=\sum_x|\psi^x_{\scriptscriptstyle B}\rangle\langle x_{\scriptscriptstyle B}|,
\end{align}
clearly, $U^{B\dagger}U^{B}=U^BU^{B\dagger}=I^B$.
Then we can construct a controlled unitary in $\mathcal{H}_A\otimes\mathcal{H}_B$,
\begin{align}
U^{AB}=|1_{\scriptscriptstyle A}\rangle\langle 1_{\scriptscriptstyle A}|\otimes U^B+\mathbb{I}_1^A\otimes \mathbb{I}^B,
\end{align}
where $\mathbb{I}_1^A$ is the identity operator in the residual subspace of $H^A$. Obviously, $U^{AB}$ is an unitary operator in $H_A\otimes H_B$. Denote
\begin{align}
V=&\sum_n(|1_{\scriptscriptstyle A}\rangle\langle n_{\scriptscriptstyle A}|\otimes|n_{\scriptscriptstyle B}\rangle\langle 1_{\scriptscriptstyle B}|+|n_{\scriptscriptstyle A}\rangle\langle 1_{\scriptscriptstyle A}|\otimes|1_{\scriptscriptstyle B}\rangle\langle n_{\scriptscriptstyle B}|)\nonumber\\
-&|1_{\scriptscriptstyle A}\rangle\langle1_{\scriptscriptstyle A}|\otimes|1_{\scriptscriptstyle B}\rangle\langle1_{\scriptscriptstyle B}|,
\end{align}
and
\begin{align}
W=V+\mathbb{I}_2,
\end{align}
where $\mathbb{I}_2$ is the identity operator from the residual subspace  which  eliminates $VV^\dagger$ in $H^A\otimes H^B$, and $W$ is an unitary operator swapping specified states. It is easy to see that
\begin{align}
E_x(\theta)=\langle x_{\scriptscriptstyle B}|U^{AB}W|1_{\scriptscriptstyle B}\rangle,
\end{align}
based on Stinespring dilation theorem, $\{E_x(\theta)\}$ could be implement by an unitary $U^{AB}W$ on $\rho\otimes|1_{\scriptscriptstyle B}\rangle\langle1_{\scriptscriptstyle B}|$ and projective measurement $\{|x_{\scriptscriptstyle B}\rangle\langle x_{\scriptscriptstyle B}|\}$.

\bibliography{ref}

\end{document}